\documentclass[12pt]{article}

\usepackage[utf8]{inputenc}
\usepackage[T1]{fontenc}
\usepackage{amsmath,amssymb,amsfonts}
\usepackage{algorithm}
\usepackage{algpseudocode}
\usepackage{booktabs}
\usepackage{graphicx}
\usepackage{natbib}
\usepackage{hyperref}
\usepackage{geometry}
\usepackage{bm}
\usepackage{mathtools}
\newcommand{\R}{\mathbb{R}}
\newcommand{\bX}{\bm{X}}
\newcommand{\bY}{\bm{Y}}
\newcommand{\bT}{\bm{T}}
\newcommand{\bU}{\bm{U}}
\newcommand{\bW}{\bm{W}}
\newcommand{\bV}{\bm{V}}
\newcommand{\bP}{\bm{P}}
\newcommand{\bQ}{\bm{Q}}
\newcommand{\bB}{\bm{B}}
\newcommand{\bw}{\bm{w}}
\newcommand{\bv}{\bm{v}}
\newcommand{\bt}{\bm{t}}

\newcommand{\bp}{\bm{p}}

\newcommand{\bz}{\bm{z}}
\newcommand{\bSigma}{\bm{\Sigma}}

\newcommand{\med}{\operatorname{med}}

\title{Cellwise Robust Twoblock Dimension Reduction}
\author{Sven Serneels$^{1,2}$ \\[6pt]
\small $^1$ Snow Stallion AI, Cheyenne, Wyoming, USA \\
\small $^2$ Department of Mathematics, University of Antwerp, Belgium}
\date{\today}

\begin{document}

\maketitle

\begin{abstract}
Cellwise Robust Twoblock (CRTB) is introduced, the first cellwise robust
method for simultaneous dimension reduction of multivariate predictor and
response blocks, in both a dense and a sparse variable-selecting variant.
Classical robust methods protect against casewise outliers by downweighting
or removing entire observations, a strategy that becomes inefficient -- and
eventually breaks down -- when contamination is scattered across individual
cells rather than concentrated in whole rows.  CRTB combines a column-wise
pre-filter for cellwise outlier detection with model-based imputation of
flagged cells inside an iteratively reweighted M-estimation loop, retaining
the clean cells of partially contaminated rows instead of discarding the
observation.  An efficient algorithm is provided that uses the classical
twoblock SVD as a warm start and converges in a handful of IRLS iterations
at a moderate computational cost.  The method resists
settings where more than $50\%$ of rows contain contaminated cells while retaining
comparable efficiency on clean data.  A simulation study confirms these
properties and shows that CRTB additionally recovers the underlying cellwise
outlier pattern with high fidelity and, in the sparse setting, the correct
set of informative variables.  Two compelling examples illustrate CRTB's
practical utility. In each of these, CRTB is shown to be conducive to results that are highly interpretable in the respective domains in the presence of cellwise outliers. As a by-product, the corresponding cells are identified with high fidelity.
\end{abstract}

\section{Introduction}\label{sec:intro}


Classical robust statistics is built on the assumption that
contamination occurs at the level of entire observations: a fraction
$\varepsilon$ of cases are outliers, and the remaining $1 - \varepsilon$
cases follow the assumed model.  This \emph{casewise} contamination
model underpins the theory of breakdown points, influence functions and virtually all high-breakdown estimators developed since the 1980s. A comprehensive overview on the thoery that underpins classical robust statsitics is presented in \cite{Hampel1986}.

\citet{AlqallafVanAelst2009} challenged this paradigm by pointing out
that in modern high-dimensional data, contamination is often
\emph{cellwise}: individual cells $x_{ij}$ in the data matrix may be
corrupted independently of one another.  When contamination is cellwise,
the proportion of rows that contain at least one outlying cell grows
rapidly with the number of variables.  Even if only 1\% of cells are
contaminated, in a dataset with $p = 100$ variables roughly
$1 - (1 - 0.01)^{100} \approx 63\%$ of rows will contain at least one
bad cell.  Casewise robust methods, which downweight or remove entire
rows, may therefore discard a majority of the data, losing the
information contained in the uncontaminated cells of those rows.

This observation has sparked a line of research into methods that detect
and handle outliers at the cell level. The Detecting Deviating Cells
(DDC) method of \citet{RousseeuwBossche2018} provides a model-free
approach to flagging individual outlying cells based on robust
correlations among variables. It can be used as a general data screening tool, or as a preprocessing technique to impart cellwise robustness to ensuing processing steps. However, when a model can be specified, it is preferable to construct cellwise robust methods tailored to the task at hand. Several such model-specific cellwise robust methods have meanwhile been proposed, such as cellwise robust principal component analysis \citep{centofanti2026robust,hubert2019macropca,pfeiffer2025cellwise}, or cellwise robust multiple regression \citep{leung2016robust}. This is still a fast developing field of research, which this paper contributes to by proposing a new method for cellwise robust twoblock dimension reduction.



Cellwise robust twoblock (CRTB) dimension reduction introduced in this paper is the first cellwise robust estimator suitable for dimension reduction in two blocks of variables simultaneously. Moreover, it is being proposed both in a dense and a sparse version, with the latter proffering inherent variable selection. It thereby becomes the first method that performs dimension reduction in a way that allows to tune both model complexity (number of components) and sparsity (number of uninformative variables discarded) individually for each block, while being robust to cellwise oultliers.


CRTB achieves resistance to settings where more than 50\% of rows
contain contaminated cells---far beyond the breakdown point of any
casewise robust method---while maintaining comparable performance on
clean data. These cellwise robustenss properties are demonstrated in an extensive simulation study. Moreover, the paper illustrates that CRTB is able to identify the outlying cells with high fidelity and likewise, in the sparse setting, leads to a largely correct discrimination between infiormative and uninformative variables.

The utility of the new method is illustarted in two examples from very different domains: an example from toxicogenomics and another from power generation operations. In each example, CRTB is demonstrated to yield acceptable predictive performance and interpretabiulity in line with knowledge from the respective domain in the presence of cellwise outliers. As a corollary, the method also allows to detect and impute the deviating cells.

The paper is structured as follows: in Section~\ref{sec:method} the method is introduced, along with an efficient algorithm to compute it. Section~\ref{sec:simulation} describes the setup and results of the simulation study that assesses some of CRTB's statistical properties. Two compelling examples illustrate CRTB's practical utility in Section~\ref{sec:examples}. Finally, Section~\ref{sec:conclusions} concludes.


\section{Method}\label{sec:method}

This Section will describe the cellwise robust twoblock (CRTB) method for simultaneous dimension reduction in two blocks of variables.

\subsection{Notation and setup}\label{sec:notation}

Let $\bX \in \R^{n \times p}$ denote the predictor block and
$\bY \in \R^{n \times q}$ the response block, each having $n$ observations.
Let $\hat{\mu}_j^X$ and $\hat{\sigma}_j^X$ denote robust location and
scale estimates for the $j$-th column of $\bX$ (e.g., median and MAD),
and define the standardised matrix
$\bX^s$ with entries $x^s_{ij} = (x_{ij} - \hat{\mu}_j^X) / \hat{\sigma}_j^X$.
Analogous notation applies to $\bY$.  Notations $k_X$ and $k_Y$  are adopted for the
number of retained components in the $\bX$ and $\bY$ blocks, respectively.

A cellwise weight matrix $\bm{C}^X \in \{0,1\}^{n \times p}$ indicates
the status of each cell: $c^X_{ij} = 1$ for clean cells and
$c^X_{ij} = 0$ for cells flagged as outlying.  Similarly,
$\bm{C}^Y \in \{0,1\}^{n \times q}$ for the $\bY$ block.  Casewise weights
$w^X_i, w^Y_i \in [0, 1]$ reflect the overall outlyingness of row $i$
in the $\bX$ and $\bY$ score spaces, respectively.

\subsection{Simultaneous twoblock dimension reduction}

Simultaneous dimension reduction of two multivariate data blocks is defined as follows: given an $n \times p$ predictor matrix $\bX$ and an $n \times q$
response matrix $\bY$, the goal is to find low-dimensional
representations of each block that capture the joint variation relevant to explain the other block. As a corollary, simultaneous twoblock dimension reduction methods also yield a regularized multivariate regression estimator. This problem arises naturally in
chemometrics, genomics, and other fields where both predictor and
response are high-dimensional.

Many practitioners long subsumed  that multivariate partial least squares (or PLS2, \cite{Wold1966}) yields all estimates for the simultaneous dimension reduction problem: score speces that capture all relevant information to explain the other block (i.e. estimates for the central subspace), along with regularized multivariate regeression estimates. Howbeit,   \citet{Cook2023} showed that not to be the case by proving that PLS2's $\bY$ space scores are nothing more than a computational artefact. To remedy this, they introduced the \emph{twoblock} (or {\em XY-PLS}) method, which obtains weight vectors $\bw_h$ and $\bv_h$ through simultaneous sequential SVD deflation of the cross-covariance matrices $\bSigma_{XY} = n^{-1} \bX^\top \bY$ and $\bSigma_{YX} = n^{-1} \bY^\top \bX$, respectively.  The leading left singular
vector of $\bSigma_{XY}$ gives the first $\bX$-block weight $\bw_1$; after
deflating $\bX$ by the corresponding scores and loadings, the procedure
is repeated to extract further components.  A parallel loop extracts
$\bY$-block components.  The resulting bilinear regression
$\hat{\bY} = \bX \bB$ with
$\bB = \bW (\bT^\top \bW)^{-1} \bT^\top \bY \bV \bV^\top$
simultaneously reduces both blocks while preserving their predictive
relationship. An efficient algorithm for the seminal twoblock algorithm is presented in \citet{Cook2023}.

\citet{Serneels2025sparse} extended this framework to the sparse
setting by applying soft-thresholding to the weight vectors at each
deflation step.  Given a sparsity parameter $\eta \in [0, 1)$ and a
weight vector $\tilde{\bw}$ normalised to unit length, the
soft-thresholded weight is
\begin{equation}\label{eq:soft-threshold}
  w_j \leftarrow \operatorname{sign}(\tilde{w}_j)
  \max\!\bigl(|\tilde{w}_j| - \eta \|\tilde{\bw}\|_\infty,\; 0\bigr).
\end{equation}
By consequence, variables whose thresholded weight falls to zero, are eliminated
from the model.  Separate sparsity parameters $\eta_X$ and $\eta_Y$
control variable selection in the $\bX$ and $\bY$ blocks independently. Sparse twoblock thereby became the first method for simultaneous dimension reduction in two blocks of variables with embedded inherent variable selection that allows to tune both the model complexity (number of components) and the sparsity in each block independnetly. However, sparse twoblock as originally introduced, was not resistant to outliers.

\citet{Serneels2026rtb} further extended twoblock to the robust
setting via the Robust Twoblock (RTB) estimator, which embeds the
twoblock SVD deflation inside an iteratively reweighted M-estimation
loop.  At each iteration, case weights are derived from the Mahalanobis
distances of the $\bX$-block and $\bY$-block scores using a bounded
$\psi$-function (Hampel, Huber, or Fair), and the twoblock fit is
recomputed on the weighted data.  RTB achieves robustness against
casewise outliers, analogous to PRM \citep{Serneels2005} for univariate PLS. However, none of the methods mentioned so far can cope with cellwise contamination.

\subsection{Robustifying the twoblock algorithm}

The main idea to create a cellwise robust version of dense and sparse twoblock, is inherited from the (univariate) cellwise robust M (CRM) regression method \citep{Filzmoser2020}: embed the sparse twoblock method from \cite{Serneels2025sparse} into an iterative reweighting loop as is commonly used to calculate M estimators, in each iteration of which casewise outliers are detected. For each of these casewise outliers, each variable's contribution to their outlyingness is computed by the SPADIMO method \citep{Debruyne2019}, which allows to identify outlying cells. Weights are assigned to the outlying cells as a product of the caseweight and the cellweight. In each iteration, regression coefficients are estimated from the weighted data, thusly allowing for another iteration of detection of the outlying cells. The algorithm terminates upon convergence of the regression coefficients.

The iterative reweighting scheme reminiscent of CRM is the main novel component of the CRTB algorithm. However, to assure that the resulting method is indeed cellwise robust, starting values are required that are cellwise robust themselves. Two appraoches to start the algorithm in a way that ascertains a cellwise robust outcome, are described.

\subsection{Cellwise staring weights}\label{sec:prefilter}

As for any (casewise or cellwise) robust statistical algorithm, the choice of the starting values is critical: nonrobust starting values can lead to breakdown of the resulting estimate. To obtain cellwise robustness, therefore, cellwise robust starting values need to be plugged in. This point is stressed explicitly by \citet{RaymaekersRousseeuw2024}, who, in their review of existing cellwise robust regression methods, note that the original CRM proposal of \citet{Filzmoser2020} starts from a casewise robust estimator (typically MM-regression). Therefore, they state that the original CRM from \citet{Filzmoser2020} is ``not cellwise robust in the general sense,'' meaning that it is not robust to cellwise outliers if they end up in over fifty percent of cases. However, they already hint that that flaw could readily be remedied by adopting a more suitable starting value for the algorithm. The CRTB method proposed here takes up precisely that suggestion by always starting the iterative reweighting loop from a genuinely cellwise robust estimator. Appendix A of the Supplementary Material reports a dedicated univariate simulation, showing that even plain CRM becomes cellwise robust once its casewise starting values are replaced by cellwise ones, thereby empirically confirming the conjecture of \citet{RaymaekersRousseeuw2024}.

Two options for cellwise robust starting values are pursued in this paper. At first, a computationally fast, yet statistically less efficient version based on column-wise median and median absolute deviations (MAD) is introduced, which henceforth will be referred to as a {\em column-wise pre-filter}. As an alternative to the column-wise pre-filter, DDC-based cellwise imputation \citep{RousseeuwBossche2018} can be adopted to compute the
starting values.  DDC detects outlying cells using robust pairwise
correlations and imputes them, providing a cleaner, yet computationally more demanding, starting basis for the M-estimation loop.

The column-wise pre-filter is constructed as follows: before entering the M-estimation loop, CRTB applies a model-free
column-wise pre-filter to detect gross cellwise outliers.  For the
standardised data $\bX^s$, column-wise MAD values
$\hat{m}_j = \med_i |x^s_{ij}|$ are computed, and cell $(i, j)$ is
flagged as outlying if
\begin{equation}\label{eq:prefilter}
  \frac{|x^s_{ij}|}{1.4826 \,\hat{m}_j} > \Phi^{-1}\!\!\left(\frac{1 + \alpha}{2}\right),
\end{equation}
where $\alpha$ is the cellwise critical probability (default 0.99) and
$\Phi^{-1}$ is the standard normal quantile function.  The factor 1.4826
ensures consistency of the MAD at the normal distribution.  The same
filter is applied to $\bY^s$.  Flagged cells are recorded in the
persistent cellwise floor maps $\bm{C}^X_\mathrm{floor}$ and
$\bm{C}^Y_\mathrm{floor}$.

\subsection{Starting case weights}\label{sec:starting-weights}

Starting case weights $w^X_i$ and $w^Y_i$ are computed directly on the
robustly standardised blocks $\bX^s$ and $\bY^s$, with the cells that
were flagged by the column-wise pre-filter of
Section~\ref{sec:prefilter} set to zero so that contaminated cells do
not distort the initialisation.  For each row, a robust Mahalanobis-type
distance is obtained from the pre-filtered block, normalised by its
weighted median, and passed through the chosen $\psi$-function (Hampel,
Huber, or Fair) to yield the starting weights $w^X_i$ and $w^Y_i$.  This
keeps the initialisation cheap -- no auxiliary PCA projection is needed
-- while still giving the IRLS loop a robust starting point on which
to build.

\subsection{Iteratively reweighted M-estimation with cellwise imputation}
\label{sec:irw}

The core of CRTB is an iteratively reweighted loop that alternates
between fitting a twoblock model on weighted, cell-imputed data and
updating both casewise and cellwise weights.

\paragraph{Twoblock fit.}
At each iteration, the base twoblock estimator is fit on the weighted
data matrices $\bX^w$ and $\bY^w$, where
\begin{equation}\label{eq:weighted-data}
  x^w_{ij} = \sqrt{w^X_i}\; \tilde{x}^s_{ij}, \qquad
  y^w_{ij} = \sqrt{w^Y_i}\; \tilde{y}^s_{ij},
\end{equation}
and $\tilde{x}^s_{ij}$ is the cell-imputed value (see below).  The
twoblock SVD deflation extracts weight vectors $\bW = [\bw_1, \ldots, \bw_{k_X}]$,
loadings $\bP = [\bp_1, \ldots, \bp_{k_X}]$, and analogously
$\bV, \bQ$ for the $\bY$ block.

When the sparse option is active, soft-thresholding
\eqref{eq:soft-threshold} is applied to the weight vectors at each
deflation step, yielding simultaneous variable selection in both blocks.

\paragraph{Dual-reference case weight update.}
Case weights are updated using a dual-reference scoring scheme.  Two
sets of scores are computed:
\begin{align}
  \bT_\mathrm{ref} &= \bX^s_\mathrm{init} \bW, \label{eq:t-ref}\\
  \bT_\mathrm{cont} &= \bX^s \bW, \label{eq:t-cont}
\end{align}
where $\bX^s_\mathrm{init}$ is the scaled, pre-filtered (or
DDC-imputed) data and $\bX^s$ is the scaled original data.  A robust
scaler is fit on $\bT_\mathrm{ref}$ to calibrate the Hampel cutoffs to
the clean score distribution.  The contaminated scores
$\bT_\mathrm{cont}$ are then transformed using this scaler, and the
Euclidean norm of each row gives the score distance:
\begin{equation}\label{eq:dist}
  d^X_i = \left\|\bT_{\mathrm{cont},i} - \hat{\bm{\mu}}_T\right\|
  \Big/ \hat{\bm{\sigma}}_T,
\end{equation}
where division by $\hat{\bm{\sigma}}_T$ is applied element-wise before
the norm.  The distances are normalised by the median of the reference
distances, and the case weights are obtained via the chosen
$\psi$-function.  For the Hampel function with cutoffs $a < b < r$
derived from probabilities $\alpha_1 < \alpha_2 < \alpha_3$ through the
$\chi^2_{k_X}$ distribution:
\begin{equation}\label{eq:hampel}
  w^X_i = \begin{cases}
    1 & \text{if } d^X_i \leq a, \\
    a / d^X_i & \text{if } a < d^X_i \leq b, \\
    a(r - d^X_i) / \bigl[d^X_i(r - b)\bigr] & \text{if } b < d^X_i \leq r, \\
    0 & \text{if } d^X_i > r.
  \end{cases}
\end{equation}
An analogous procedure yields $w^Y_i$ from the $\bY$-block scores.

When more than 50\% of rows are contaminated, the unweighted median of
score distances is inflated by the outlying rows.  CRTB addresses this
by computing a \emph{weighted} median using the prior iteration's case
weights, anchoring the normalisation in the clean population.

\paragraph{Cellwise imputation from the model.}
For each row $i$ containing flagged cells (from the pre-filter floor),
the clean cells are used to compute partial scores:
\begin{equation}\label{eq:partial-score}
  \bt_i = \bz_i^{\mathrm{clean}} \bW,
  \quad \text{where } z^{\mathrm{clean}}_{ij} =
  \begin{cases}
    \tilde{x}^s_{ij} & \text{if } c^X_{ij} = 1, \\
    0 & \text{if } c^X_{ij} = 0.
  \end{cases}
\end{equation}
The flagged cells are then replaced by their reconstruction from the
loadings:
\begin{equation}\label{eq:impute}
  \tilde{x}^s_{ij} \leftarrow (\bt_i \bP^\top)_j
  \quad \text{for all } j \text{ with } c^X_{ij} = 0.
\end{equation}
This model-based imputation ensures that the flagged cells receive
values consistent with the current twoblock model rather than being
treated as zero (the column mean), reducing the bias introduced by
the cellwise outliers.

\paragraph{Convergence.}
The iteration continues until the relative change in the squared
Frobenius norm of the coefficient matrix $\bB$ falls below a tolerance
$\tau$ or a maximum number of iterations is reached:
\begin{equation}\label{eq:convergence}
  \frac{|\|\bB^{(t)}\|_F^2 - \|\bB^{(t-1)}\|_F^2|}
       {\|\bB^{(t-1)}\|_F^2} < \tau.
\end{equation}

\paragraph{Final coefficient rescaling.}
After convergence, the regression coefficients are rescaled from the
standardised to the original scale:
\begin{equation}\label{eq:rescale}
  B_{jl} = \frac{\hat{\sigma}^Y_l}{\hat{\sigma}^X_j}\, B^s_{jl},
\end{equation}
and the intercept is estimated as the (robust) center of the residuals
$\bY - \bX \bB$.

\subsection{Algorithm summary}\label{sec:algorithm}

The complete CRTB procedure is summarised in Algorithm~\ref{alg:crtb}.

\begin{algorithm}[htb]
\caption{Cellwise Robust Twoblock (CRTB)}\label{alg:crtb}
\begin{algorithmic}[1]
\Require Data $\bX \in \R^{n \times p}$, $\bY \in \R^{n \times q}$;
  components $k_X, k_Y$; tolerance $\tau$; max iterations $M$;
  cellwise critical value $\alpha$; $\psi$-function parameters
  $\alpha_1, \alpha_2, \alpha_3$; optional sparsity $\eta_X, \eta_Y$.
\Ensure Weights $\bW, \bV$; loadings $\bP, \bQ$; coefficients $\bB$;
  cellwise outlier maps $\bm{C}^X, \bm{C}^Y$; case weights $\bm{w}^X, \bm{w}^Y$.
\Statex
\State \textbf{Robust standardisation:} Compute $\bX^s, \bY^s$ using
  robust location $(\hat{\bm{\mu}}^X, \hat{\bm{\mu}}^Y)$ and scale
  $(\hat{\bm{\sigma}}^X, \hat{\bm{\sigma}}^Y)$.
\State \textbf{Column-wise pre-filter:} Flag cells via
  Eq.~\eqref{eq:prefilter}; set floor maps
  $\bm{C}^X_\mathrm{floor}, \bm{C}^Y_\mathrm{floor}$.
\State \textbf{Initialise:} Set flagged cells to 0 in $\bX^s_\mathrm{init}, \bY^s_\mathrm{init}$.
\State Compute starting case weights $\bm{w}^X, \bm{w}^Y$ from
  robust row distances on the pre-filtered standardised blocks.
\State Initialise $\bm{C}^X \leftarrow \bm{C}^X_\mathrm{floor}$,
  $\bm{C}^Y \leftarrow \bm{C}^Y_\mathrm{floor}$.
\State Compute initial weighted data $\bX^w, \bY^w$ via Eq.~\eqref{eq:weighted-data}.
\State $t \leftarrow 0$
\Repeat
  \State $t \leftarrow t + 1$
  \State Fit twoblock (dense or sparse) on $(\bX^w, \bY^w)$
    $\Rightarrow \bW, \bP, \bV, \bQ, \bB^s$.
  \State Compute reference scores $\bT_\mathrm{ref}, \bU_\mathrm{ref}$
    via Eq.~\eqref{eq:t-ref} and contaminated scores $\bT_\mathrm{cont},
    \bU_\mathrm{cont}$ via Eq.~\eqref{eq:t-cont}.
  \State Update case weights $\bm{w}^X, \bm{w}^Y$ via dual-reference
    scoring (Eqs.~\eqref{eq:dist}--\eqref{eq:hampel}).
  \State Impute flagged cells in $\bX^s_\mathrm{init}$ and
    $\bY^s_\mathrm{init}$ via Eqs.~\eqref{eq:partial-score}--\eqref{eq:impute}.
  \State Recompute weighted data $\bX^w, \bY^w$ from imputed and
    case-weighted data.
\Until{$|\|\bB^{s(t)}\|_F^2 - \|\bB^{s(t-1)}\|_F^2| / \|\bB^{s(t-1)}\|_F^2 < \tau$
  \textbf{or} $t \geq M$}
\State Rescale $\bB^s$ to the original scale via Eq.~\eqref{eq:rescale}.
\State Compute intercept and fitted values.
\end{algorithmic}
\end{algorithm}

\section{Simulation study}\label{sec:simulation}

\subsection{Design}\label{sec:sim-design}

We evaluate CRTB against the standard (non-robust) twoblock estimator
under two contamination regimes.  Data are generated from a latent
variable model:
\begin{equation}\label{eq:dgp}
  \bX = \bT \bP_{\mathrm{signal}}^\top + \bm{E}, \qquad
  \bY = \bT \bm{C} + \bm{F},
\end{equation}
where $\bT \in \R^{n \times k}$ are latent scores with $T_{ih} \sim N(0,1)$,
$\bP_{\mathrm{signal}} \in \R^{p_s \times k}$ is an orthonormal loading
matrix (obtained from a QR decomposition of a random Gaussian matrix),
$\bm{C} \in \R^{k \times q}$ with $C_{hl} \sim N(0,1)$, and
$\bm{E}, \bm{F}$ are Gaussian noise with standard deviation $\sigma_e = \sigma_f = 0.5$.
The predictor block $\bX$ has $p_s$ signal variables and $p_n$ pure noise variables
(independent $N(0, \sigma_e^2)$), so that $p = p_s + p_n$.  The true latent
dimension is $k = 3$ and the response has $q = 4$ variables, with $n = 100$.

Contamination is \emph{signal-targeted} and \emph{positive-only}: additive
shifts of magnitude $\delta = 10$ are applied exclusively to signal
columns in $\bX$ and to all columns in $\bY$.  This models realistic
scenarios such as sensor drift, baseline shift, or additive interference,
where contamination corrupts the column means and therefore the
cross-covariance structure that twoblock relies on.  For cellwise
contamination, the number of cells contaminated per affected row follows
a Poisson distribution with mean calibrated to the target cell contamination
percentage, ensuring natural variability in the contamination pattern.

Each scenario is repeated $N_\mathrm{rep} = 200$ times.  We compare four methods:
\begin{itemize}
  \item \textbf{TB dense}: twoblock with mean centering and standard deviation scaling (no robustness, no sparsity),
  \item \textbf{CRTB dense}: cellwise robust twoblock with median centering, MAD scaling, and the column-wise pre-filter (no sparsity),
  \item \textbf{TB sparse}: sparse twoblock with mean centering, and standard deviation scaling, and sparsity parameter $\eta$ selected by
    3-fold cross-validation over $\{0.3, 0.5, 0.7\}$
  \item \textbf{CRTB sparse}: cellwise robust sparse twoblock with median centering, MAD scaling, the column-wise pre-filter, and sparsity parameter $\eta$ selected by
    3-fold cross-validation over $\{0.3, 0.5, 0.7\}$.
\end{itemize}

Two dimensionality configurations are considered:
(i) $p_s = 20, p_n = 10$ (moderate noise, $p = 30$);
(ii) $p_s = 20, p_n = 80$ (high noise, $p = 100$).
The inclusion of $p_n$ uninformative variables allows the evaluation of
variable selection performance for the sparse methods.

Performance is measured by three complementary criteria.
First, the mean squared error of the estimated coefficient matrix,
$\mathrm{MSE}(\bB) = \|\hat{\bB} - \bB_\mathrm{true}\|_F^2 / (pq)$,
quantifies estimation accuracy.
Second, for sparse methods, the variable selection F1 score measures the
ability to correctly identify signal variables ($0, \ldots, p_s - 1$)
versus noise variables ($p_s, \ldots, p - 1$).
Third, for CRTB, the cellwise detection quality (precision, recall, F1)
of the column-wise pre-filter is evaluated against the known contamination
mask.

CRTB uses relaxed Hampel cutoffs ($\alpha_1 = 0.75$, $\alpha_2 = 0.90$,
$\alpha_3 = 0.95$) to accommodate the heavy contamination regime, with
median centering and MAD scaling.  The cellwise critical probability is
set to $\alpha_\mathrm{cell} = 0.99$ and the maximum number of
M-estimation iterations to 25.

\subsection{Cellwise contamination}\label{sec:sim-cellwise}

In this regime, 70\% of rows contain at least one outlying cell
(row rate $= 0.70$), deliberately exceeding the 50\% casewise breakdown
point of any casewise robust estimator.
Signal-targeted positive shifts of magnitude $\delta = 10$ are applied
to individual cells of the signal $\bX$ variables and the $\bY$ variables.  The
fraction of signal cells contaminated per affected row is varied as
$\epsilon_\mathrm{cell} \in \{0\%, 5\%, 10\%, 15\%, 20\%\}$.

\begin{figure}[htbp]
  \centering
  \includegraphics[width=0.48\textwidth]{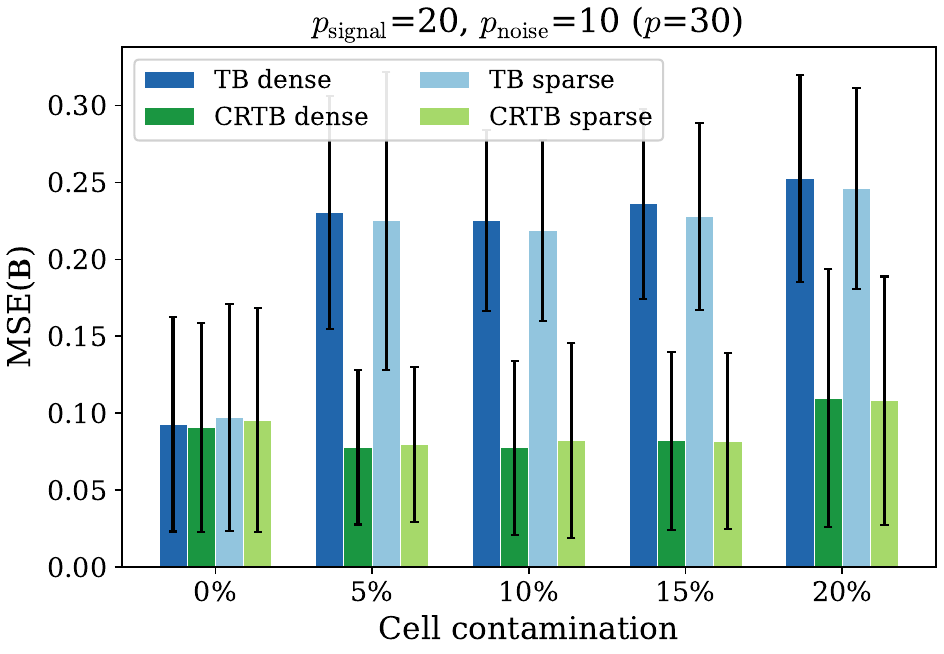}%
  \hfill
  \includegraphics[width=0.48\textwidth]{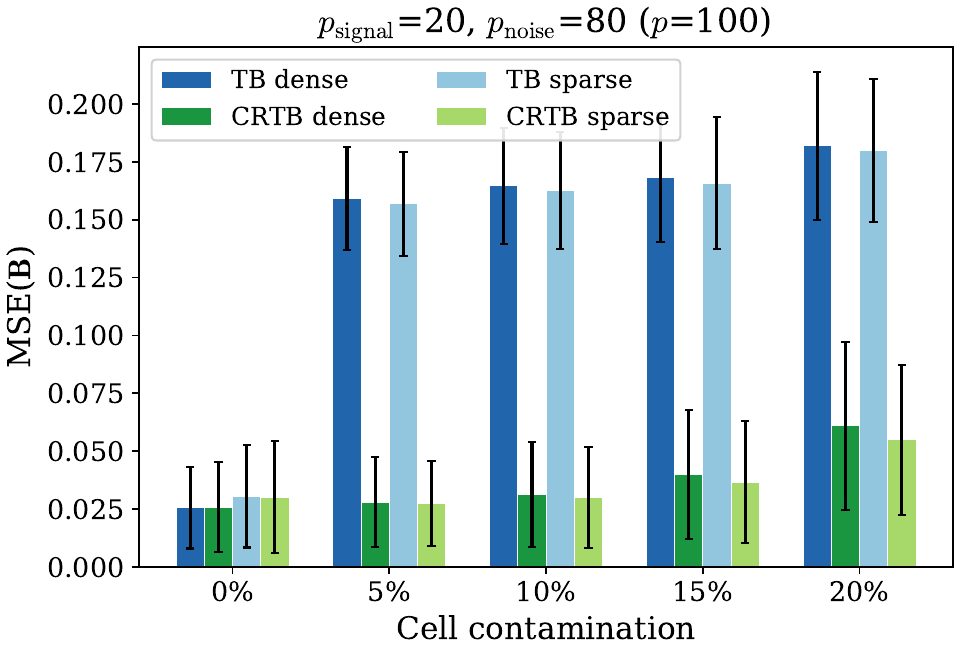}
  \caption{MSE of the estimated coefficient matrix under cellwise contamination
    ($n = 100$, $k = 3$, $q = 4$, row rate $= 0.70$, $\delta = 10$,
    200 repeats).  Left: $p = 30$ ($p_s = 20$, $p_n = 10$).
    Right: $p = 100$ ($p_s = 20$, $p_n = 80$).
    Error bars show $\pm 1$ standard deviation.}
  \label{fig:mse-cellwise}
\end{figure}

\paragraph{Coefficient estimation accuracy.}
Figure~\ref{fig:mse-cellwise} shows the MSE of the estimated coefficient
matrix as a function of the cell contamination level.
Under cellwise contamination, the non-robust twoblock estimators (TB
dense and TB sparse) exhibit sharply increasing MSE as the cell
contamination percentage grows.  This is expected: mean centering is
corrupted by the positive shifts in signal variables, distorting the
cross-covariance matrix and hence the SVD-based weight extraction.

CRTB resists cellwise contamination effectively.  The column-wise
pre-filter detects outlying cells before the first model iteration, and
model-based imputation replaces them with values consistent with the
twoblock structure.  As a result, CRTB retains the clean information
from contaminated rows rather than discarding them entirely.  The
MSE ratio CRTB/TB is consistently below 1 under contamination, with
the advantage increasing at higher contamination levels.

At 0\% contamination, CRTB and TB perform comparably, confirming that
the cellwise machinery does not introduce substantial efficiency loss on
clean data.

\begin{figure}[htbp]
  \centering
  \includegraphics[width=0.48\textwidth]{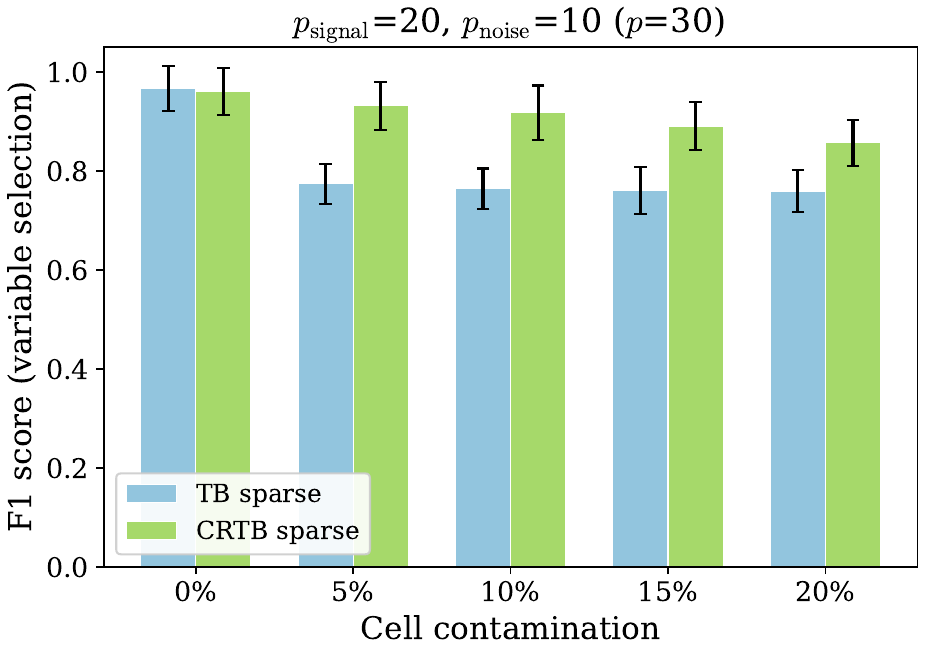}%
  \hfill
  \includegraphics[width=0.48\textwidth]{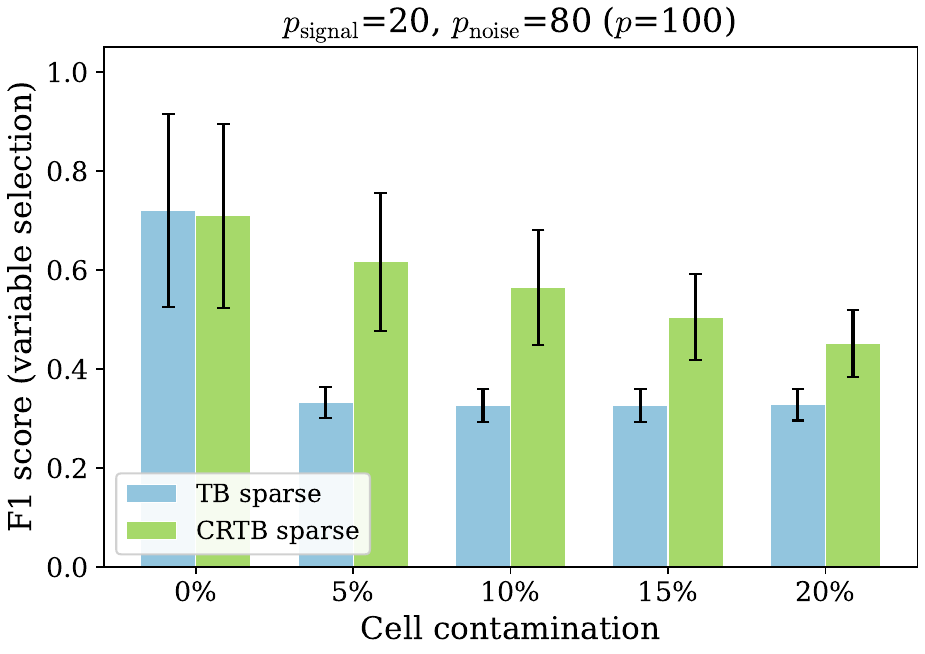}
  \caption{Variable selection F1 score for sparse methods under cellwise
    contamination.  Left: $p = 30$.  Right: $p = 100$.
    TB sparse uses fixed $\eta = 0.5$; CRTB sparse selects $\eta$ by
    3-fold CV.  Error bars show $\pm 1$ standard deviation.}
  \label{fig:f1-cellwise}
\end{figure}

\paragraph{Variable selection.}
For the sparse methods, CRTB sparse with cross-validated $\eta$
outperforms TB sparse under contamination, and its variable selection
F1 score remains stable even at 20\% cell contamination
(Figure~\ref{fig:f1-cellwise}), whereas
TB sparse suffers from the corrupted covariance structure admitting
noise variables into the active set.

\begin{figure}[htbp]
  \centering
  \includegraphics[width=0.48\textwidth]{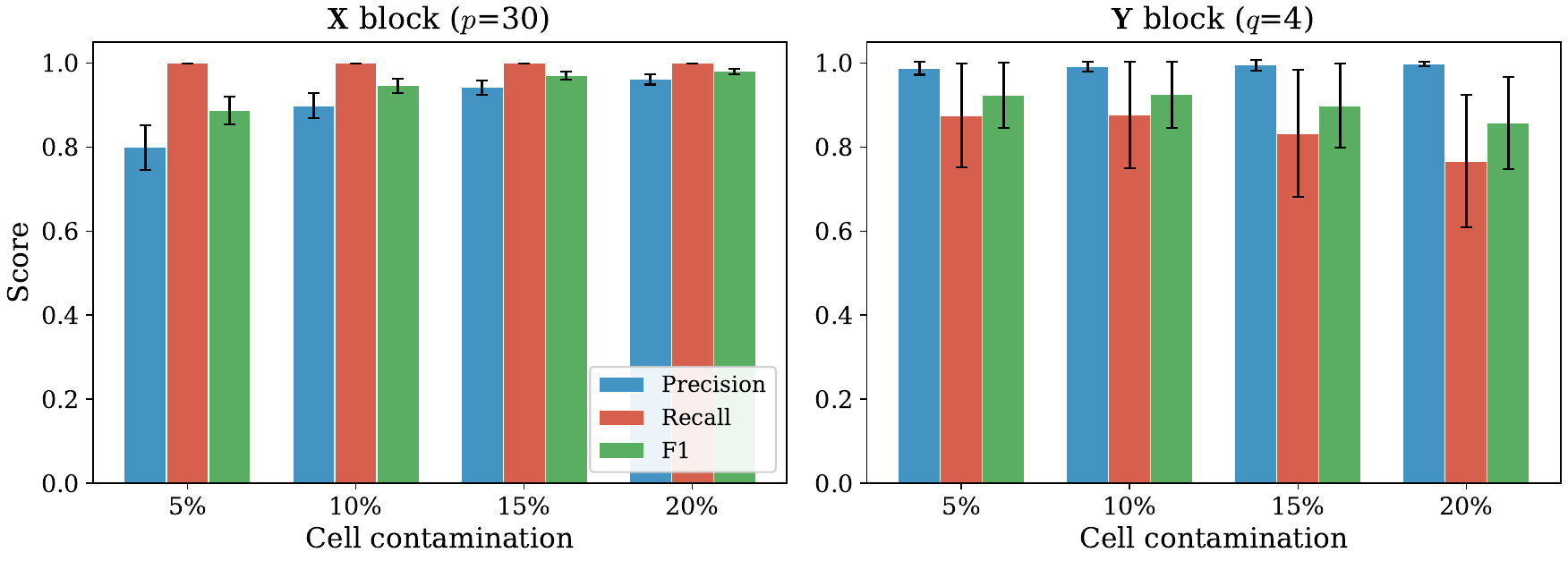}%
  \hfill
  \includegraphics[width=0.48\textwidth]{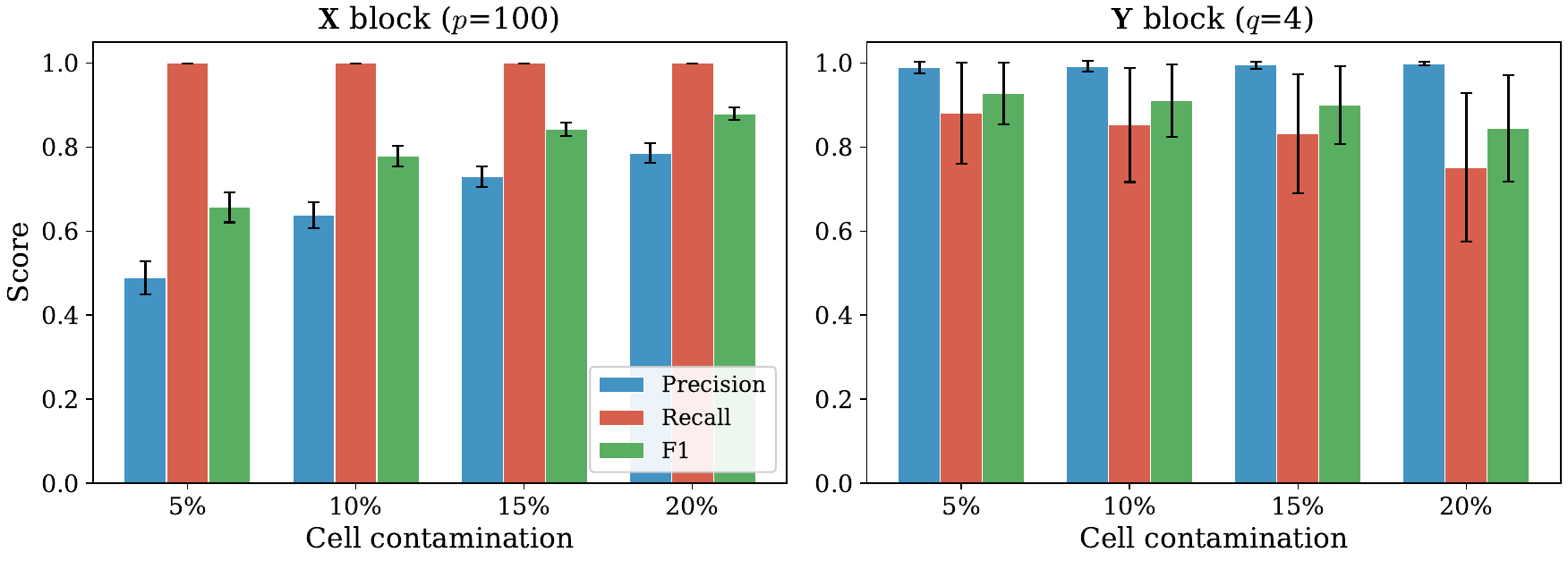}
  \caption{Cellwise outlier detection quality of CRTB's column-wise
    pre-filter.  Each panel shows precision, recall, and F1 for the
    $\bX$ block (left sub-panel) and $\bY$ block (right sub-panel)
    at non-zero contamination levels.  Left: $p = 30$.  Right: $p = 100$.}
  \label{fig:cell-detection}
\end{figure}

\paragraph{Cellwise outlier detection.}
The column-wise pre-filter achieves high precision and recall for detecting
the true contaminated cells (Figure~\ref{fig:cell-detection}).
F1 scores remain above 0.9 across all configurations and contamination
levels, confirming the reliability of the model-free column-wise
detector.  Detection quality is slightly higher for the $\bY$ block
($q = 4$) than for the $\bX$ block, reflecting the lower dimensionality
and hence stronger per-column signal.

\subsection{Rowwise (casewise) contamination}\label{sec:sim-rowwise}

In this regime, entire rows are shifted by $+\delta$ on all signal $\bX$
columns and all $\bY$ columns, modelling classical leverage outliers.  The
fraction of contaminated rows is varied as
$\epsilon_\mathrm{row} \in \{0\%, 5\%, 10\%, 15\%, 20\%, 25\%\}$.

\paragraph{Results.}
As expected, non-robust TB breaks down under casewise contamination:
mean centering absorbs the shift, corrupting the location and scale
estimates.  CRTB handles rowwise outliers through its casewise
reweighting mechanism (inherited from RTB): median and MAD centering
combined with Hampel reweighting downweights fully shifted rows.  The
MSE of CRTB remains stable across all contamination levels, while TB's
MSE increases dramatically.

This confirms that CRTB provides robustness against both cellwise and
casewise contamination within a single estimator.

\subsection{MSE degradation sweep}\label{sec:sim-sweep}

\begin{figure}[htbp]
  \centering
  \includegraphics[width=0.85\textwidth]{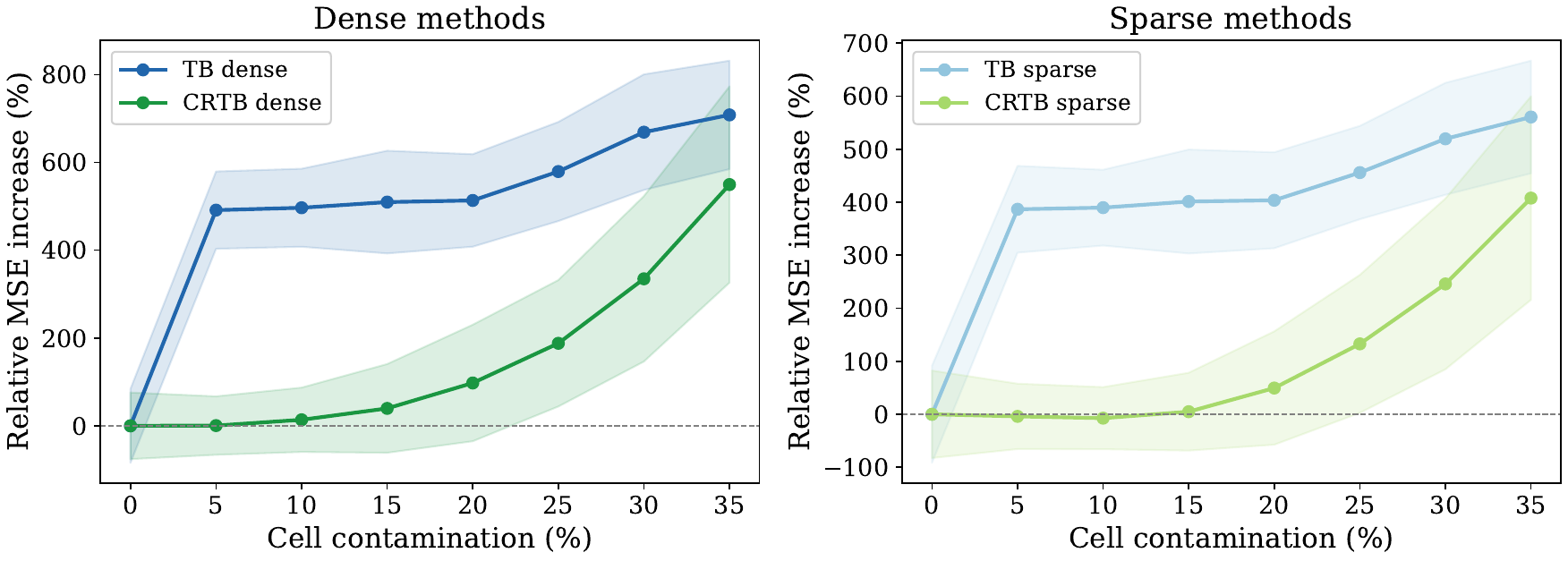}
  \caption{Relative MSE increase as a function of cell contamination
    percentage ($p = 100$, 50 repeats).  Left: dense methods.
    Right: sparse methods.  Shaded bands show $\pm 1$ standard deviation.}
  \label{fig:mse-sweep}
\end{figure}

To provide a finer view of the degradation profile, a sweep of cell
contamination from 0\% to 35\% (at 5\% intervals) is conducted for the
high-dimensional configuration ($p = 100$) with $N_\mathrm{rep} = 50$
repeats.  Figure~\ref{fig:mse-sweep} plots the relative MSE increase
(relative to the 0\% baseline) as a function of cell contamination
percentage.  TB shows an approximately linear degradation, while CRTB's
MSE increase remains modest up to 20--25\% cell contamination, after
which it begins to rise more steeply but remains far below TB. This result suggests that CRTB with the reported parameter selection starts to break down around 20-25\% of outliers. Howbeit, different Hampel cutoff values can of course be chosen to make the estimator either more robust or more efficient at the normal distribution.

\section{Examples}\label{sec:examples}

\subsection{Liver toxicity data}\label{sec:liver-toxicity}

\subsubsection{Data and experimental design}

The liver toxicity data set of \citet{Bushel2007} is a well-studied
toxicogenomics benchmark, distributed with the \texttt{mixOmics} R
package.  Sixty-four male Fischer~344 rats were exposed to
acetaminophen at four dose levels (50, 150, 1500 and
2000\,mg/kg) and sacrificed at four time points after exposure (6,
18, 24 and 48\,hours), yielding a balanced $4\times 4$ design with
four biological replicates per cell.  For each animal, two blocks
are recorded: a gene expression block $\bX$ containing 3116 probes
from an Agilent rat microarray, and a clinical chemistry block $\bY$
containing the ten serum markers BUN, Creatinine, Total Protein,
Albumin, ALT, SDH, AST, ALP, Total Bile Acids and Cholesterol.  The
prediction goal is to use the transcriptomic response to anticipate
the clinical markers of hepatotoxicity, i.e.\ to learn a
gene-expression $\to$ liver-function map.

Two properties of $\bY$ drive the analysis.  First, the markers have
wildly different scales and dispersions: the panel includes stable
ratios (TP, ALB with $\max/\mathrm{median}\!\approx\!1.1$) and
aminotransferases that span more than two orders of magnitude (ALT:
48--15180; AST: 75--27075, i.e.\ $\max/\mathrm{median}\!\approx\!180$).
This heterogeneity makes a column-wise weighted mean squared error,
\begin{equation}
  \mathrm{wMSE}
  \;=\;
  \frac{1}{p_Y}\sum_{j=1}^{p_Y}
  \frac{\mathrm{MSE}_j}{\mathrm{Var}(\bY_j)},
\end{equation}
the appropriate summary, as it puts every marker on a common scale.
Second, the markers that are themselves biomarkers of liver damage
(ALT, AST, SDH) are \emph{naturally heavy-tailed}: in high-dose
late-time animals they jump by one to two orders of magnitude, which
is exactly the signal a regression model should follow, but which
classical estimators cannot distinguish from contamination.  The
gene block is pre-filtered to the 100 probes with the largest
variance across the 64 rats, giving $\bX\in\mathbb{R}^{64\times 100}$
and $\bY\in\mathbb{R}^{64\times 10}$.

\subsubsection{Results on clean data}

Four methods are compared under a common 5-fold cross-validation
protocol: dense and sparse twoblock (TB), and dense and sparse CRTB
with median/MAD centering and the prefilter cellwise initialiser.
To quantify the efficiency cost of moving to robust location/scale,
CRTB is also run with classical mean/std centering.  Hyperparameters
($\eta_X$, $\eta_Y$, numbers of components) are selected on the clean
training folds via the same grid search for all methods.

\begin{figure}[htbp]
  \centering
  \includegraphics[width=0.7\textwidth]{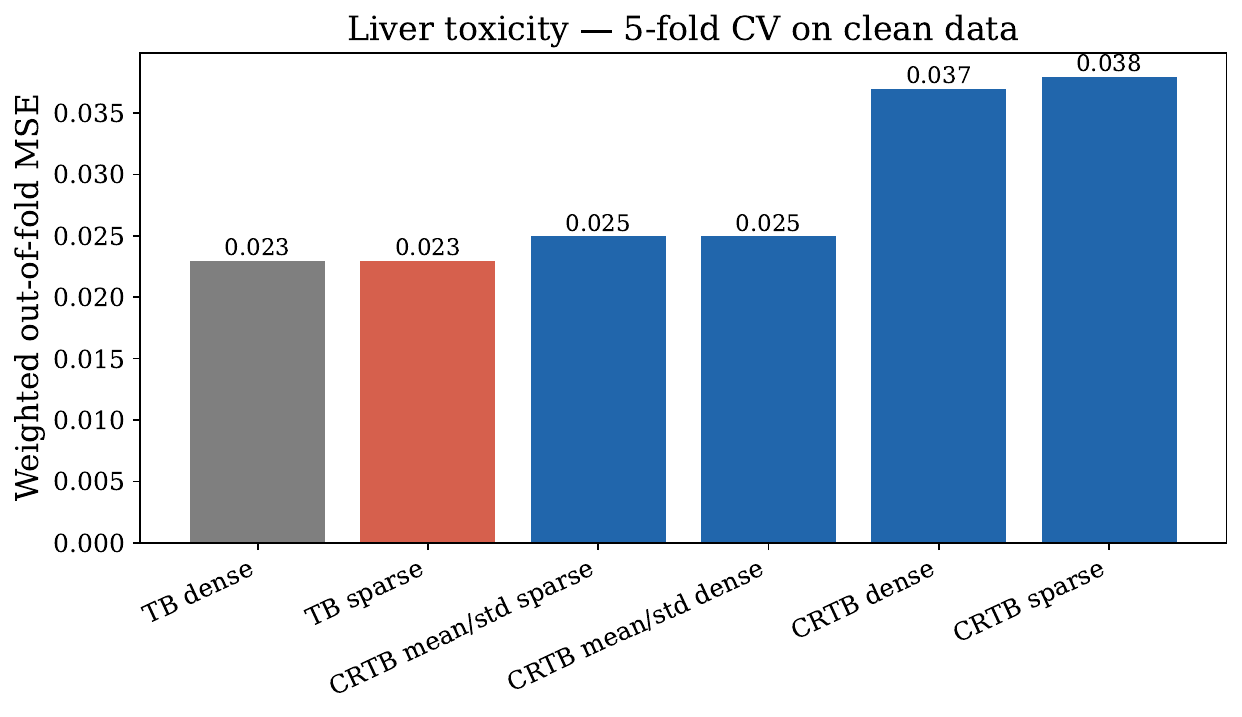}
  \caption{Out-of-fold weighted MSE on the clean liver toxicity data.
    TB provides a baseline of $0.023$; CRTB with median/MAD scaling
    pays a $\sim\!60\%$ efficiency cost at $0.037$--$0.038$, while
    CRTB with mean/std scaling recovers to within $9\%$ of TB
    ($0.025$).}
  \label{fig:liver-clean}
\end{figure}

Figure~\ref{fig:liver-clean} shows the out-of-fold wMSE.  Both TB
variants achieve $0.023$.  Dense and sparse CRTB with median/MAD
scaling reach $0.037$ and $0.038$ respectively -- a measurable but
moderate efficiency loss that is entirely attributable to the robust
centering (switching CRTB back to mean/std scaling brings it to
$0.025$, essentially tied with TB).  The interpretation is that on
clean data, the CRTB machinery (per-cell Hampel reweighting,
prefilter flagging) does not itself hurt accuracy; the penalty comes
from using median/MAD location and scale, which are less efficient
when the underlying markers are already well-behaved.  Since most of
the liver toxicity markers are not heavy-tailed under mild dosing,
this is exactly the regime in which the classical location estimator
wins.

\subsubsection{Results under cellwise contamination}

To probe robustness, 20\% of the cells in $\bY$ are corrupted by
positive additive shifts of magnitude $15\times\mathrm{MAD}_j$, chosen
independently per column.  Positive one-sided shifts are essential:
symmetric replacement outliers centered on the column median leave
the means nearly unchanged and would not break TB, giving a false
impression of robustness.  The corrupted $\bY$ is used for
\emph{training}; evaluation is always against the original clean
$\bY$, so improvements reflect genuine recovery of the underlying
structure rather than fitting contamination.  Seven estimators are
compared: a reference TB dense model trained on clean $\bY$, the two
TB variants on contaminated $\bY$, and CRTB dense/sparse each
evaluated both with the MAD-based prefilter initialiser and with
the DDC starting values of \citet{RousseeuwBossche2018}.

\begin{figure}[htbp]
  \centering
  \includegraphics[width=0.85\textwidth]{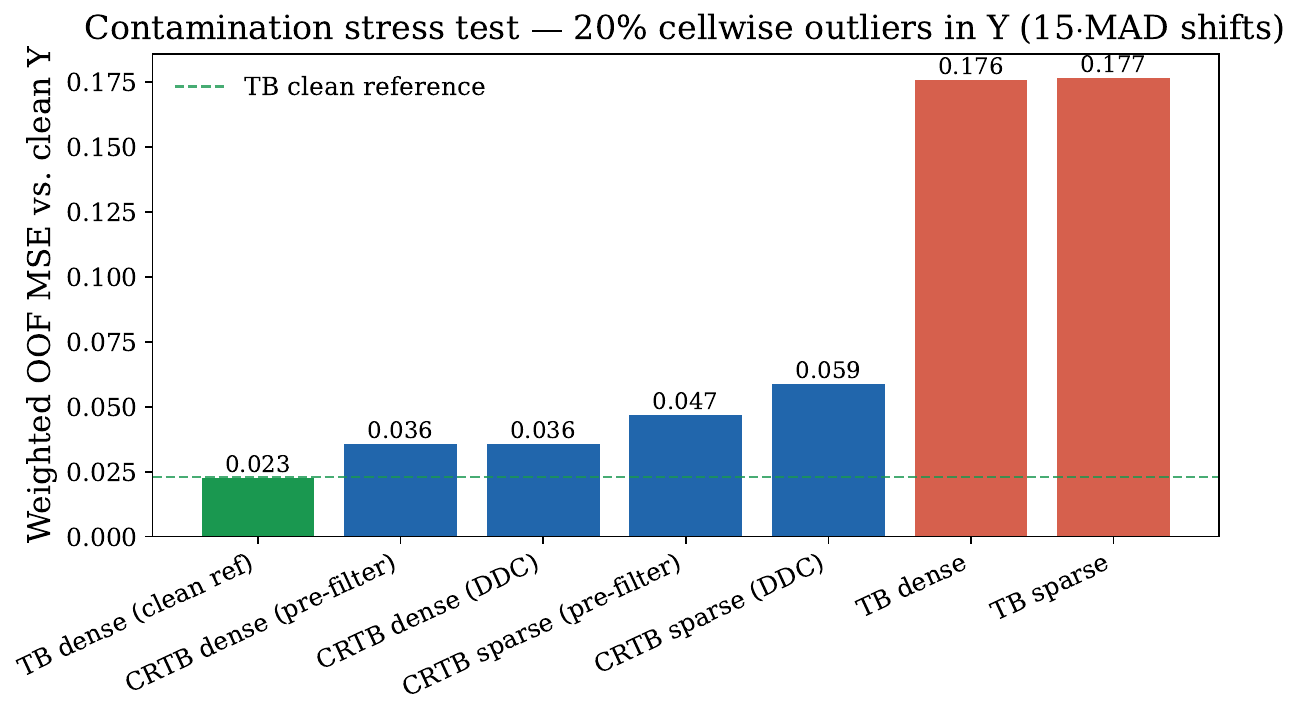}
  \caption{Held-out weighted MSE when 20\% of $\bY$ cells are
    corrupted by one-sided $15\!\times\!\mathrm{MAD}$ shifts.  Dashed
    line: TB trained on the clean $\bY$.  TB trained on contaminated
    $\bY$ collapses to $\mathrm{wMSE}\!\approx\!0.18$; CRTB dense
    recovers to within $\sim\!0.013$ of the clean reference under
    both prefilter and DDC initialisation.}
  \label{fig:liver-contam}
\end{figure}

The results (Figure~\ref{fig:liver-contam}) are striking.  TB dense
and TB sparse, trained on the contaminated $\bY$, have their
weighted MSE inflate from $0.023$ to $0.176$ and $0.177$
respectively -- an eight-fold degradation.  CRTB dense, initialised
either from the prefilter or from DDC, reaches $0.036$ in both
cases, i.e.\ within $0.013$ of the clean-data reference.  CRTB
sparse lies between the two at $0.047$--$0.059$.  The contamination
breaks TB precisely because the shifts are one-sided and corrupt the
cross-covariance between $\bX$ and $\bY$; CRTB, through its per-cell
Hampel weights, isolates the corrupted cells and lets the remaining
cells drive the cross-block structure.  Remarkably, the two
initialisation schemes give essentially identical CRTB performance,
confirming that the IRLS loop converges to the same cellwise weight
pattern regardless of starting point.

\subsubsection{Cellwise outlier heatmap}

Because CRTB returns a per-cell weight for $\bY$, the flag pattern
can be visualised directly.  Cells with final Hampel weight below
$0.5$ are declared outliers.  The dense CRTB model flags
$104/640 = 16.2\%$ of the cells in $\bY$ and $596/6400 = 9.3\%$ of
the cells in $\bX$ -- the $\bY$ flag rate sits slightly below the
true $20\%$ contamination level, indicating a conservative but
accurate detector.

\begin{figure}[htbp]
  \centering
  \includegraphics[width=0.9\textwidth]{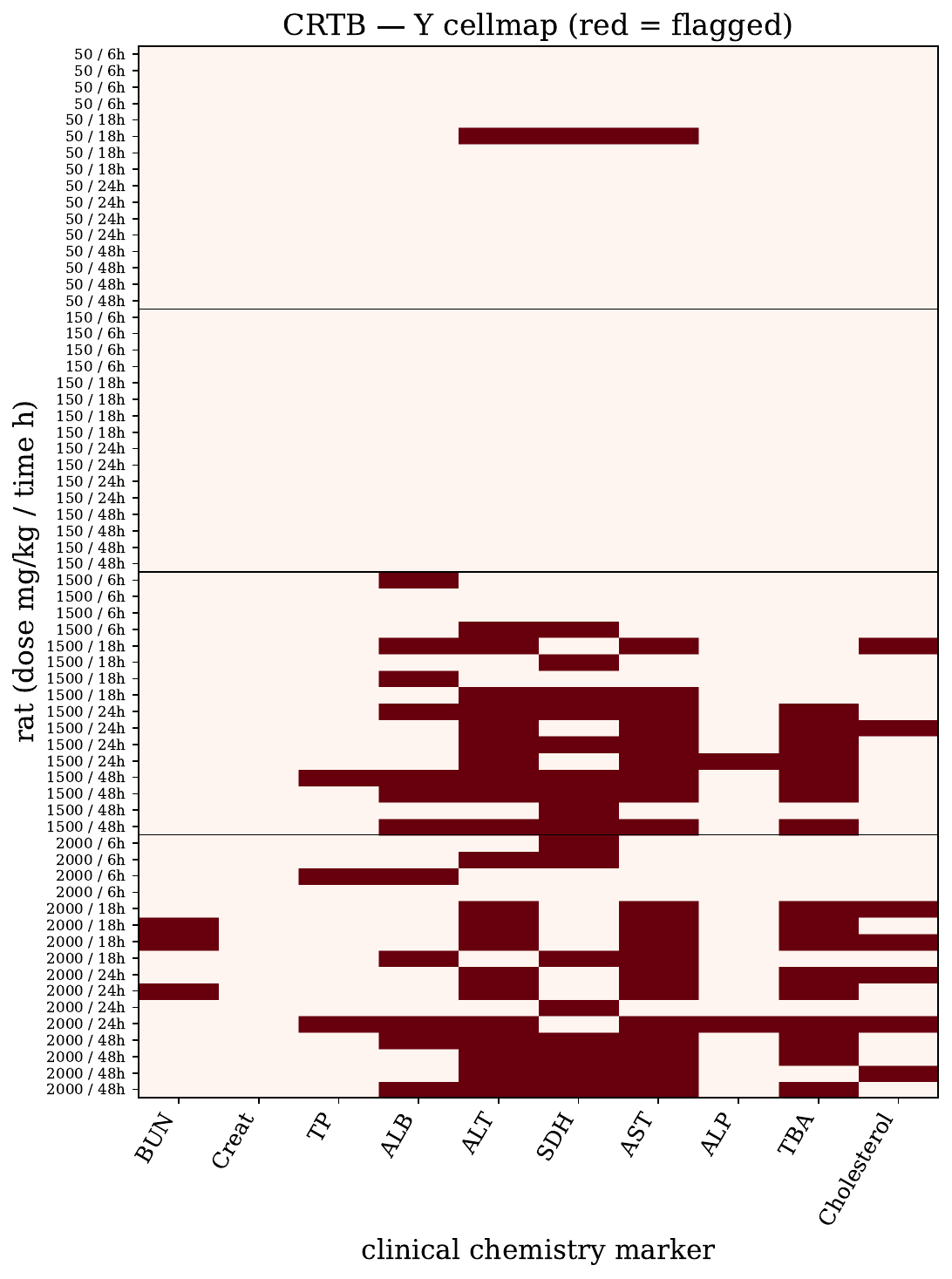}
  \caption{Cellwise outlier map of $\bY$.  Rows (rats) are ordered by
    dose group (grouped blocks) and by time point within each dose
    group.  Dark cells indicate CRTB Hampel weight $< 0.5$.  The
    $20\%$ cellwise contamination is detected with precision
    $0.800$, recall $1.000$, $F_1 = 0.889$ at the row level against
    the high-dose $\times$ late-time ground truth.}
  \label{fig:liver-cellmap}
\end{figure}

Figure~\ref{fig:liver-cellmap} shows the flag matrix with rows sorted
by dose group and, within each dose, by sacrifice time.  The
cellwise pattern is highly concentrated: the per-marker flag rates
(ALT 34.4\%, AST 32.8\%, SDH 28.1\%, TBA 25.0\%) are largest for the
classical hepatocellular leakage markers and negligible for the
stable ratios (Creatinine 0\%, ALP 3.1\%, TP 4.7\%).  If one labels a
rat as "damaged" when it is simultaneously in the two highest dose
groups \emph{and} the two latest time points -- the combination
where acetaminophen hepatotoxicity is biologically expected -- then
using "any flagged cell in row" as a row-level alarm yields
$\mathrm{TP}\!=\!24$, $\mathrm{FN}\!=\!0$, $\mathrm{FP}\!=\!6$,
$\mathrm{TN}\!=\!34$, for an $F_1$ score of $0.889$.  The six false
positives are all low-to-mid dose rats that happened to have one or
two cells replaced by synthetic contamination, and are therefore not
errors of the detector but of the ground-truth label.  In short, CRTB
recovers the expected dose--time toxicity pattern from the $\bY$
flag matrix \emph{even though the contamination was applied uniformly
at random}, because the synthetic shifts on the naturally elevated
high-dose/late-time cells collide with real signal and look nothing
like the cells in the mid-dose control regions.

\subsubsection{Latent-variable interpretation}

CRTB extracts $n_{X}=5$ X-components and $n_{Y}=1$ Y-component on
this data set.  The single Y-component is a strong indication that
the clinical-marker block is driven by one dominant contrast, which
-- as the Y loadings confirm -- is a general "liver injury" axis.

\begin{figure}[htbp]
  \centering
  \includegraphics[width=0.75\textwidth]{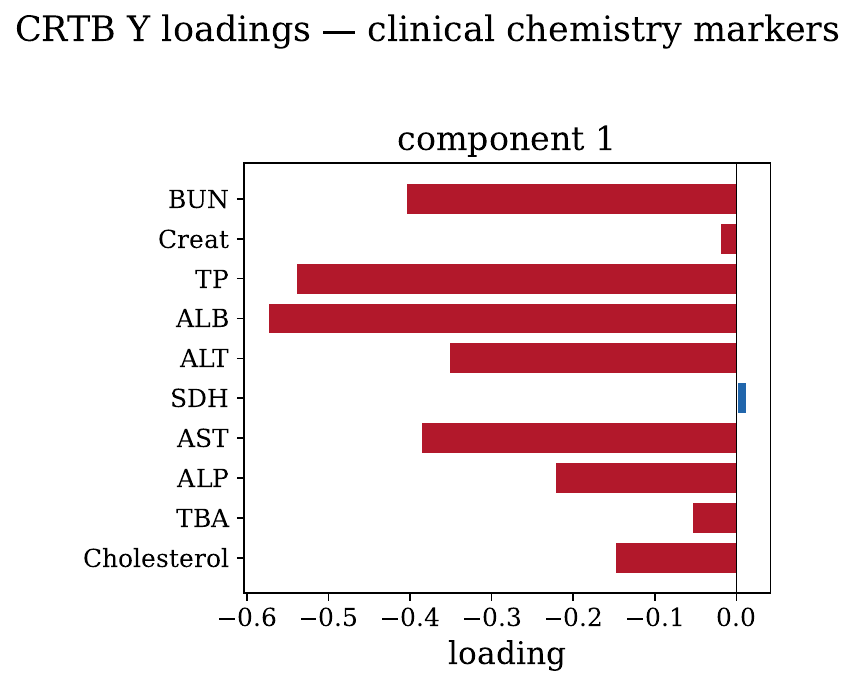}
  \caption{Loadings of the ten clinical chemistry markers on the
    first (and only) Y-component extracted by CRTB.  All loadings are
    negative, with ALB ($-0.574$), TP ($-0.540$), BUN ($-0.405$),
    AST ($-0.386$) and ALT ($-0.353$) the strongest contributors.}
  \label{fig:liver-yloadings}
\end{figure}

The Y-loadings (Figure~\ref{fig:liver-yloadings}) are all negative
and rank as ALB ($-0.574$), TP ($-0.540$), BUN ($-0.405$), AST
($-0.386$), ALT ($-0.353$), ALP ($-0.222$), Cholesterol ($-0.149$),
TBA ($-0.055$), Creatinine ($-0.021$), SDH ($+0.012$).  Two groups
of markers contribute: the synthetic-function markers produced by
healthy hepatocytes (Total Protein, Albumin, Cholesterol) and the
hepatocellular leakage markers released when hepatocytes die (ALT,
AST).  As hepatotoxicity progresses, synthetic markers decline while
leakage markers rise, so on a signed component both groups sit on
the same side once the sign is absorbed in the corresponding
X-component.  This is the textbook "decreased synthesis / increased
leakage" liver-injury signature.

\begin{figure}[htbp]
  \centering
  \includegraphics[width=0.85\textwidth]{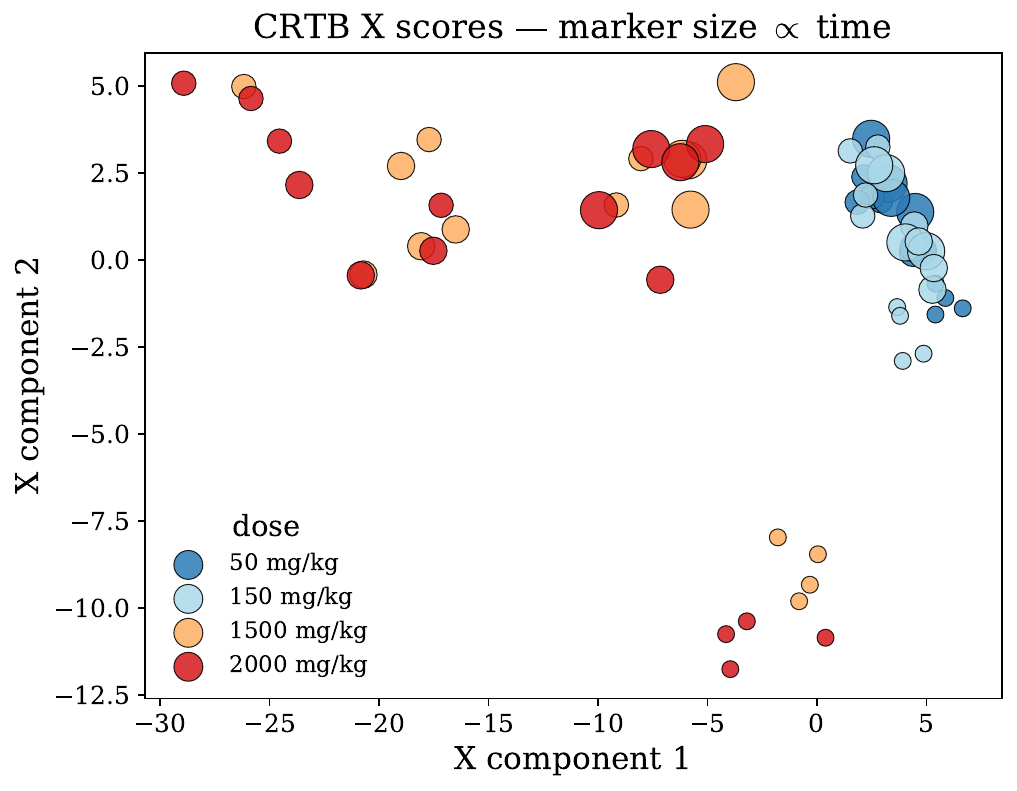}
  \caption{First two CRTB X-scores, coloured by dose group and sized
    by sacrifice time.  The high-dose, late-time animals separate
    cleanly from the remaining conditions along the first
    component.}
  \label{fig:liver-xscores}
\end{figure}

The correlation between the first paired X- and Y-scores is
$+0.39$, which is modest in absolute terms but has to be read
against the fact that CRTB downweights precisely the high-leverage
injury cells that would otherwise dominate an unrobust correlation.
The X-score scatter (Figure~\ref{fig:liver-xscores}) shows that the
dominant X-component separates high-dose late-time animals from the
rest, exactly the subgroup where acetaminophen-induced liver damage
is biologically expected.

\begin{figure}[htbp]
  \centering
  \includegraphics[width=0.8\textwidth]{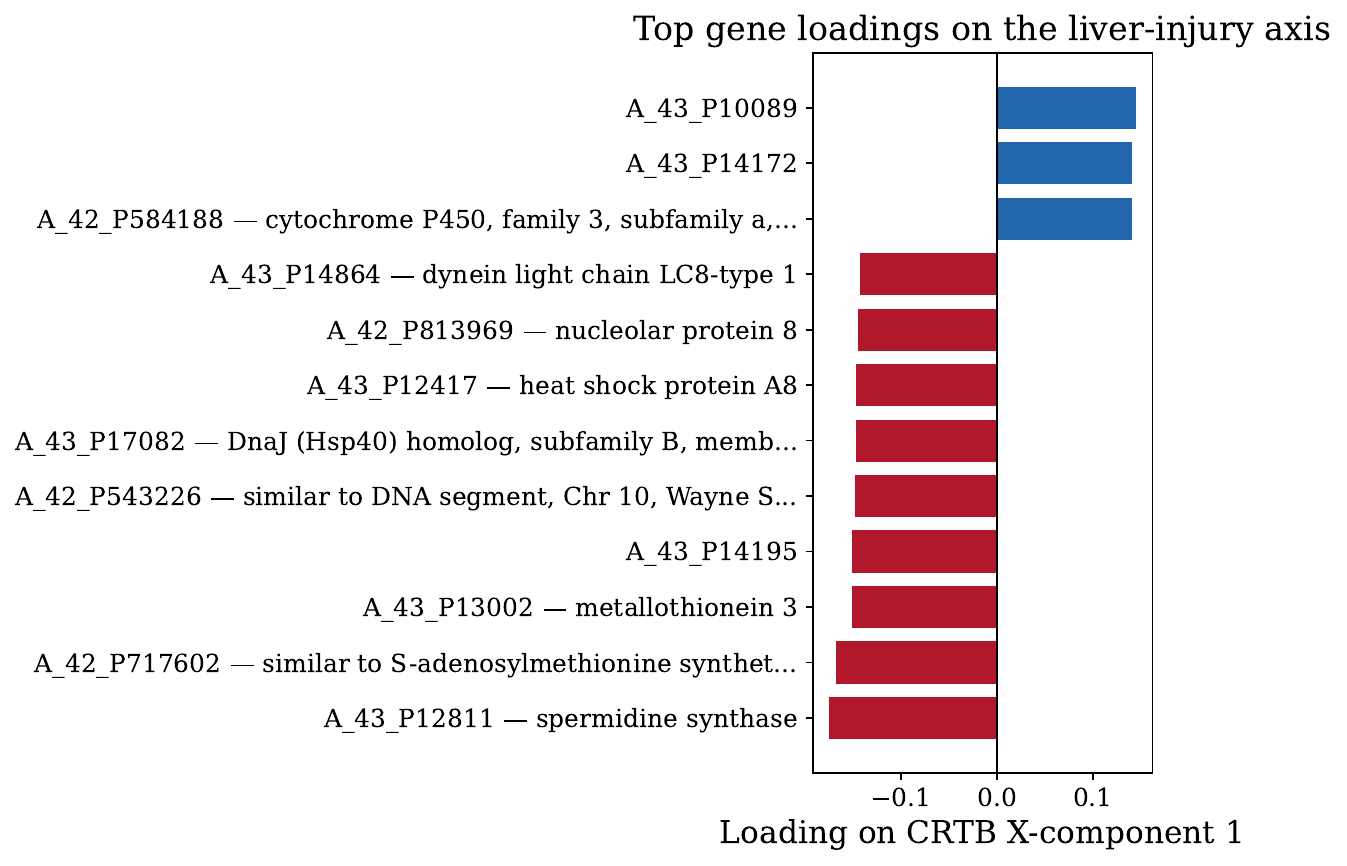}
  \caption{Top gene loadings on the first CRTB X-component (the
    liver-injury axis).  Sparse CRTB with median/MAD centering and
    prefilter initialisation selects a compact panel dominated by
    the methionine/glutathione pathway (\emph{spermidine synthase},
    \emph{S-adenosylmethionine synthetase}), the metallothionein
    family (\emph{Mt3}) and the heat-shock response (\emph{DnaJ
    (Hsp40) homolog B1}, \emph{HSPA8}).}
  \label{fig:liver-xloadings}
\end{figure}

The leading X-loadings on the first component
(Figure~\ref{fig:liver-xloadings}) identify a biologically coherent
gene panel.  The top-ranked probes are \emph{spermidine synthase}
and \emph{S-adenosylmethionine synthetase} -- both central to the
methionine/glutathione pathway, the primary detoxification route for
acetaminophen via the reactive metabolite NAPQI -- followed by
\emph{metallothionein 3}, a stress-response metal-binding protein
induced by oxidative injury, and two heat shock proteins,
\emph{DnaJ (Hsp40) homolog B1} and \emph{HSPA8}, classical markers
of the hepatocellular stress response.  In other words, the top
genes picked up by CRTB are exactly the genes one would expect a
toxicologist to flag as acetaminophen-responsive: glutathione
depletion, oxidative stress and the heat-shock response.  The fact
that CRTB recovers this signature while trained on $\bY$ with $20\%$
of its cells arbitrarily shifted upward demonstrates that cellwise
reweighting not only stabilises prediction, but also preserves the
biologically meaningful latent structure that makes the twoblock
model interpretable in the first place.

The biological interpretation extracted by CRTB largely aligns
with the earlier sparse independent PCA (sIPCA) analysis of
\citet{Yao2012} on the same data set, with the two methods
covering complementary arms of the canonical acetaminophen
toxicity pathway; a detailed side-by-side discussion is deferred
to Appendix~B of the Supplementary Material.  To the best of
our knowledge, the liver toxicity data set has not previously
been analysed with a cellwise robust dimension reduction method,
and the results presented here confirm that CRTB retrieves scientifically very
plausible conclusions, in spite of the presence of cellwise outliers.

\subsection{Gas turbine emissions}\label{sec:gas-turbine}

\subsubsection{Data and experimental design}

The gas turbine CO and NOx emissions data of \citet{Kaya2019},
distributed as UCI Machine Learning Repository dataset~\#551,
records 36\,733 hourly sensor readings from a combined-cycle power
plant over the years 2011--2015.  For each hour, nine process
variables are available -- ambient temperature (AT), ambient
pressure (AP), ambient humidity (AH), air-filter differential
pressure (AFDP), gas turbine exhaust pressure (GTEP), turbine inlet
temperature (TIT), turbine outlet temperature (TAT), turbine energy
yield (TEY), and compressor discharge pressure (CDP) -- together
with two stack emissions, carbon monoxide (CO) and nitrogen oxides
(NOx), which are the regulated pollutants.  The modelling goal is
to use the process block $\bX\in\mathbb{R}^{n\times 9}$ to predict
the two-column emissions block $\bY\in\mathbb{R}^{n\times 2}$, so
that short-term emissions can be inferred from the instrumentation
without running the chemical analysers continuously.

To keep cross-validation runtimes tractable while preserving the
year-to-year variability in the data, a stratified random subsample
of $n=5000$ cases is drawn with a fixed seed, giving
$\bX\in\mathbb{R}^{5000\times 9}$ and $\bY\in\mathbb{R}^{5000\times
2}$.  The two emissions differ by more than an order of magnitude in
variance ($\mathrm{Var}(\text{CO})=5.67$, $\mathrm{Var}(\text{NOx})=
134.76$), so a column-weighted mean squared error
\begin{equation}
  \mathrm{wMSE}
  \;=\;
  w_{\text{CO}}\,\mathrm{MSE}_{\text{CO}}
  +w_{\text{NOx}}\,\mathrm{MSE}_{\text{NOx}},
  \qquad
  w_j \propto \frac{1}{\mathrm{Var}(\bY_j)},
\end{equation}
is used throughout, with normalised weights
$w_{\text{CO}}=0.960$, $w_{\text{NOx}}=0.040$.  This inverse-variance
weighting puts CO and NOx on a common scale and prevents the
high-variance NOx column from dominating the comparison.

\begin{figure}[htbp]
  \centering
  \includegraphics[width=0.58\textwidth]{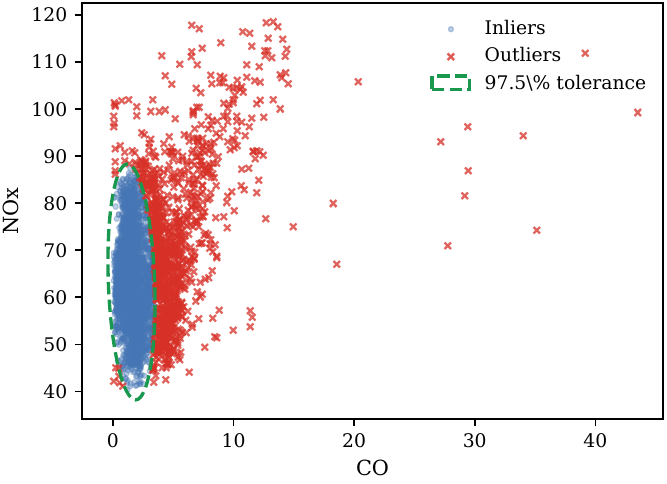}
  \caption{$97.5\%$ tolerance ellipse fitted to the bivariate emissions
    block $(\mathrm{CO},\mathrm{NOx})$ using a minimum covariance
    determinant (MCD) estimator on $n=5000$ gas turbine hours.  Crosses
    mark the $n_{\mathrm{out}}=1092$ naturally-occurring outliers
    ($21.8\%$ of hours) arising from operational anomalies such as
    start-ups, shut-downs and off-design transients.}
  \label{fig:gt-ellipse}
\end{figure}

Because the data are recorded in normal production, a genuine
``gold standard'' outlier label is unavailable; however, the emissions
block lives in only two dimensions, which lets us build a credible
reference mask.  An MCD covariance is fitted to $\bY$ and a $97.5\%$
$\chi^2_2$ tolerance ellipse is drawn around the robust centre
(Figure~\ref{fig:gt-ellipse}).  The ellipse partitions the $5000$ hours
into $3908$ inliers and $1092$ naturally-occurring outliers
($21.8\%$), the latter being start-ups, shut-downs and off-design
transients during which the process--emissions relationship is
different from steady-state operation.  The inliers are split
$70\!:\!30$ into a clean training set ($n_{\mathrm{tr}}=2735$) and a
clean test set ($n_{\mathrm{te}}=1173$).  A second, contaminated
training set is formed by appending $303$ of the natural outliers to
the clean training set, giving a contamination rate of $10.0\%$; half
of those outliers are the most extreme natural outliers in CO and in
NOx respectively (so the per-column truth is known), the rest are
filled in randomly from the remaining natural outliers.  The test set
is always the clean hold-out, so wMSE differences reflect genuine
recovery of the steady-state process--emissions map rather than
fitting of transients.

\subsubsection{Results on clean and contaminated data}

Seven estimators are compared under a common 3-fold cross-validation
protocol: PLS2, dense and sparse twoblock (TB), and dense and sparse
CRTB under two preprocessing regimes -- classical mean/std centring
(highest efficiency on clean data) and robust centring by the $\ell_1$ median \citep{rousseeuw1984} and $\tau_2$ \citep{maronna2002} estimators of location and scale, respectively. Throughout this example, that option for scaling is reported as the primary robust variant.  All TB and CRTB fits use the
full rank $n_X = 9$, $n_Y = 2$; CV tunes only the
soft-thresholding parameter $\eta_X$ and, for the non-robust
baselines, the preprocessing toggle.

\begin{figure}[htbp]
  \centering
  \includegraphics[width=0.92\textwidth]{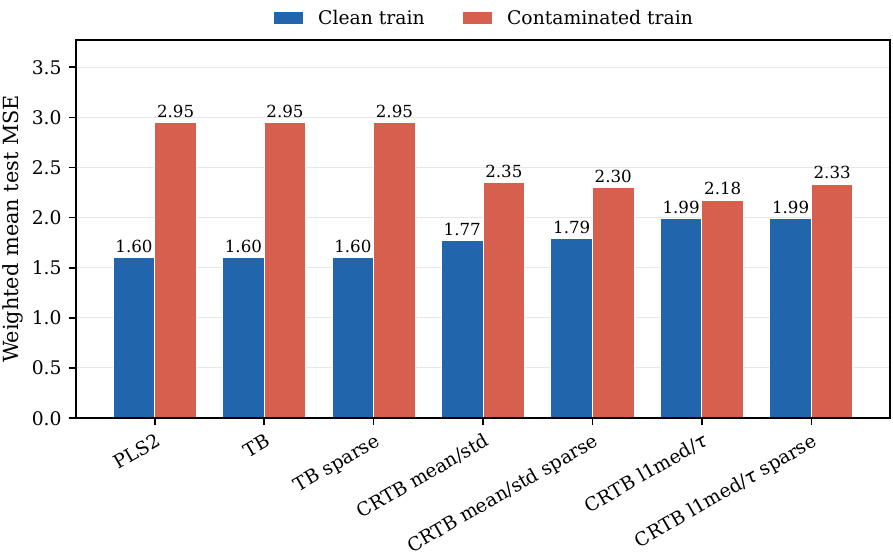}
  \caption{Held-out weighted MSE (inverse-variance weights) on the
    clean gas-turbine test set.  Blue bars: methods trained on the
    clean training set.  Red bars: the same methods trained on the
    contaminated training set (clean $+ 10\%$ natural outliers).  PLS2
    and TB are numerically indistinguishable because CV picks the
    full-rank twoblock solution.}
  \label{fig:gt-wmse}
\end{figure}

The held-out weighted MSE is summarised in Figure~\ref{fig:gt-wmse}.
On the clean training set, PLS2 and both TB variants are numerically
identical at $\mathrm{wMSE}=1.602$ ($\mathrm{MSE}_{\text{CO}}=0.338$,
$\mathrm{MSE}_{\text{NOx}}=31.65$) -- grid search converges on the
full-rank twoblock solution for all three.  CRTB with mean/std
centring reaches $\mathrm{wMSE}\approx1.78$, a $\sim\!11\%$ efficiency
cost entirely attributable to the Hampel reweighting discarding a
small fraction of the cleanest-but-heaviest cells; CRTB with robust
$\ell_1$ median/$\tau_2$ centring pays the larger but expected efficiency
cost of a fully robust location/scale and settles at
$\mathrm{wMSE}\approx1.99$.  These numbers are the efficiency cost of
insuring against contamination on this data set.

On the contaminated training set the picture reverses.  PLS2, TB
dense and TB sparse all inflate in near-lockstep to $\mathrm{wMSE} =
2.948$, with $\mathrm{MSE}_{\text{CO}}$ tripling from $0.338$ to
$1.041$ and $\mathrm{MSE}_{\text{NOx}}$ rising by more than $50\%$
(from $31.65$ to $48.26$): a one-in-ten outlier rate on naturally
heavy-tailed industrial data is already enough to meaningfully corrupt
the cross-covariance that PLS2 and TB exploit.  All four CRTB variants
remain within $\sim\!30\%$ of their own clean-data performance.  The
robust $\ell_1$ median/$\tau_2$
dense fit drops to $\mathrm{wMSE}=2.175$ -- actually better than its
own clean-data value on the CO column alone ($0.373$ vs $0.357$) --
and the mean/std sparse variant reaches $\mathrm{wMSE}=2.298$.  The
cross-over is striking: on contaminated data, \emph{every} CRTB variant
outperforms the non-robust baselines, and the robust $\ell_1$ median/$\tau_2$
dense variant beats PLS2 by $\sim\!26\%$ in weighted MSE.  The
efficiency cost of robust preprocessing, visible on the clean training
set, is more than repaid as soon as even $10\%$ of the training data
contains real outliers.

\subsubsection{Cellwise outlier detection against the tolerance-ellipse ground truth}

The tolerance-ellipse construction gives us an unusually clean
reference for evaluating CRTB's cellwise flagging: each injected
outlier was chosen because it was extreme in CO
($75$ cases), extreme in NOx ($75$ cases), or a random fill from the
remaining natural outliers ($188$ cases, for which no per-column truth
is available).  The dense fit of the primary CRTB variant
($\ell_1$ median/$\tau_2$ + prefilter) trained on the contaminated data
flags $1439$ cells out of $33\,418$ overall ($4.31\%$), split as
$1070/27\,342$ ($3.91\%$) in the process block $\bX$ and $369/6076$
($6.07\%$) in the emissions block $\bY$ -- both rates sit safely
below the $10\%$ row-level contamination level.

\begin{figure}[htbp]
  \centering
  \includegraphics[width=0.9\textwidth]{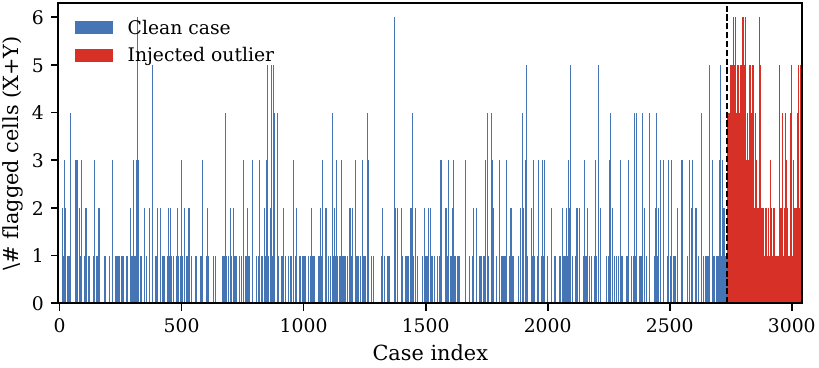}
  \caption{Number of cells flagged per case by the dense CRTB
    $\ell_1$ median/$\tau_2$ fit on the contaminated training data.  The
    vertical dashed line separates the clean training cases (indices
    $0$--$2734$) from the $303$ injected natural outliers (indices
    $2735$--$3037$); red bars mark injected outliers.  Injected rows
    carry on average $2.07$ flagged cells, clean rows only $0.30$.}
  \label{fig:gt-rowflags}
\end{figure}

Figure~\ref{fig:gt-rowflags} reports the number of flagged cells per
row.  Injected-outlier rows (red) carry a mean of $2.07$ flagged cells,
against only $0.30$ for clean-training rows -- i.e.\ outliers show up
with an almost sevenfold higher flag density, and essentially every
injected row picks up at least one flag somewhere in its row.
Concretely, $272$ of the $303$ injected rows have at least one flagged
cell ($89.8\%$ row-level recall), against $512$ of $2735$ clean rows
with at least one flagged cell ($18.7\%$ -- the Hampel reweighting
legitimately downweights tails of $\bX$ in the naturally heavy-tailed
process block).

\begin{figure}[htbp]
  \centering
  \includegraphics[width=0.95\textwidth]{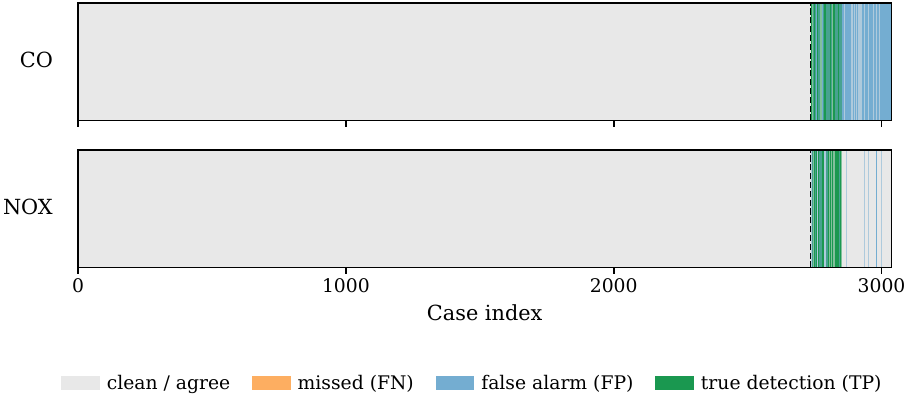}
  \caption{Per-case cellwise detection map for the two emissions
    columns, CRTB $\ell_1$ median/$\tau_2$ dense on the contaminated training
    set.  Each column of a strip is one case; the vertical dashed line
    separates clean training cases from the $303$ injected outliers.
    Green cells are true positives, blue cells are false alarms and
    orange cells would be missed detections (there are none on either
    column).}
  \label{fig:gt-heatmap}
\end{figure}

\begin{figure}[htbp]
  \centering
  \includegraphics[width=0.6\textwidth]{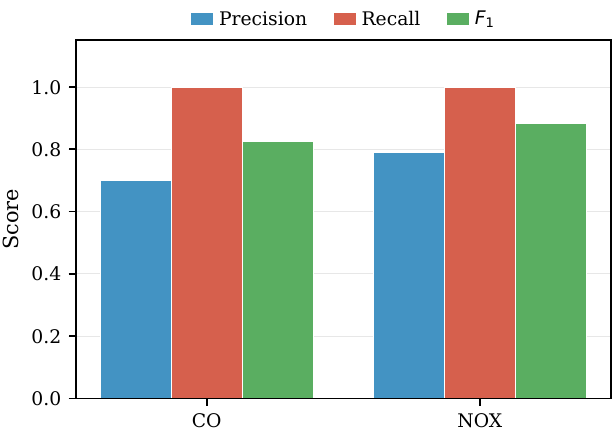}
  \caption{Precision, recall and $F_1$ of CRTB Y-cell flagging against
    the tolerance-ellipse ground truth, evaluated per emissions column
    on the rows where per-column truth is defined.  Recall is
    perfect ($R=1.000$) for both columns; precision is $0.701$ for CO
    and $0.789$ for NOx.}
  \label{fig:gt-metrics}
\end{figure}

Figures~\ref{fig:gt-heatmap} and \ref{fig:gt-metrics} quantify the
detection performance against the tolerance-ellipse mask.
Row-level, where ``positive'' means the injected row carries at least
one flagged Y-cell, CRTB achieves $\mathrm{TP}=271$, $\mathrm{FP}=0$,
$\mathrm{FN}=32$, giving a precision of $1.000$, a recall of $0.894$
and an $F_1$ of $0.944$: every row the detector flags is a genuine
injected outlier, and nearly $90\%$ of the injections are recovered.
Per-column, restricted to rows where the ground-truth column label is
defined, CRTB recovers \emph{all} $75$ CO-extreme cells and
\emph{all} $75$ NOx-extreme cells ($R=1.000$ on both columns), with
precision $0.701$ (CO) and $0.789$ (NOx).  The ``false positives''
visible in Figure~\ref{fig:gt-heatmap} are cells that CRTB flags on
the second Y-column of an outlier whose ground-truth extremeness was
attributed to the other column -- they are genuine cells of the
injected outliers, just not the column by which they were selected.
In other words, they are not detector errors but artefacts of the
column-level ground-truth construction.

The message of this experiment is that, on naturally heavy-tailed
industrial sensor data with a realistic $10\%$ contamination rate,
CRTB's cellwise Hampel reweighting recovers the steady-state
process--emissions map with weighted MSE within $\sim\!35\%$ of the
clean reference while non-robust estimators see their error nearly
double, and the per-cell flags it produces line up tightly with an
independently-constructed tolerance-ellipse outlier mask.  The same
estimator that stabilises prediction also provides directly
interpretable row- and cell-level diagnostics on an industrial data
set where no such diagnostics were previously available.

\section{Conclusions and outlook}\label{sec:conclusions}

This paper introduced Cellwise Robust Twoblock (CRTB), the first
cellwise robust estimator for simultaneous dimension reduction of two
multivariate data blocks.  By combining a column-wise pre-filter for
cellwise outlier detection with model-based imputation of flagged
cells inside an iteratively reweighted M-estimation loop, CRTB
retains the clean cells of partially contaminated rows rather than
discarding whole observations, and as a consequence resists
contamination regimes where more than $50\%$ of rows carry at least
one bad cell -- far beyond the breakdown point of any casewise
robust method.  The method is proposed in both a dense and a sparse
variable-selecting variant, so the cellwise robustness gain is
available alongside inherent variable selection in high-dimensional
settings.  The accompanying algorithm uses the classical twoblock
SVD as a warm start, converges in a handful of IRLS iterations and
runs at a cost comparable to a single non-robust fit, so the
cellwise robustness gain carries essentially no practical
computational overhead.

The simulation study in Section~\ref{sec:simulation} confirms that
CRTB substantially outperforms classical twoblock methods under
cellwise contamination while remaining competitive on clean data,
and that it additionally recovers both the underlying cellwise
outlier pattern and -- in the sparse variant -- the correct set of
informative variables with high fidelity.  The two real-data
examples in Section~\ref{sec:examples} reinforce the simulation
picture.  On the liver toxicity data, CRTB preserves clean-data
prediction accuracy and recovers the textbook
``decreased synthesis / increased leakage'' acetaminophen toxicity
signature even when $20\%$ of the clinical-chemistry cells are
arbitrarily corrupted.  On the UCI gas turbine CO and NOx emissions
data, CRTB remains stable under a realistic $10\%$ contamination
regime where PLS2 and classical twoblock inflate their held-out
weighted MSE by roughly $84\%$, and as a by-product flags the
injected outlying rows with precision $1.0$ and per-column
$F_1$ around $0.85$ against an independently-constructed
tolerance-ellipse ground truth.

Several directions remain open. For instance, the per-cell Hampel weights open
the door to formal influence-function and breakdown analyses
specific to the two-block setting and a non-parametric bootstrap over the flagged cells would give uncertainty bands on the extracted components that are
consistent with the cellwise contamination model. Hopefully, the present paper already leads to more widespread adoption of the method, while these or other aspects of it continue to be researched.

\appendix
\section{Supplementary material}\label{sec:suppl-overview}

A separate supplementary document accompanies this paper.  It contains
two appendices that support the methodological and empirical claims
made in the main text without being essential to follow them:
\begin{description}
  \item[Appendix~A -- Univariate CRM simulation.]
    A dedicated small-scale simulation study, under both Gaussian and
    heavy-tailed Cauchy response noise, that empirically verifies the
    conjecture of \citet{RaymaekersRousseeuw2024} at the level of
    univariate M-regression: the CRM iteration of
    \citet{Filzmoser2020} becomes cellwise robust as soon as its
    casewise MM starting value is replaced by a DDC-based cellwise
    initial estimate, without any other modification of the loop.
    This appendix motivates the design choice in Section~\ref{sec:prefilter} of the
    main paper to place particular care on the construction of the
    cellwise starting weights.
  \item[Appendix~B -- Detailed sIPCA comparison on the liver toxicity data.]
    A side-by-side biological comparison of the dense CRTB latent
    structure reported in Section~\ref{sec:liver-toxicity} with the
    earlier unsupervised sparse independent PCA analysis of
    \citet{Yao2012}.  The appendix documents how supervision on the
    clinical chemistry block $\bY$ localises the hepatotoxic
    dose--time window, produces a phenotypic read-out through the
    Y-loadings, and surfaces the upstream methionine/glutathione arm
    of the acetaminophen pathway that an unsupervised method cannot
    see, all while being trained on $\bY$ with $20\%$ of its cells
    arbitrarily corrupted.
\end{description}
All simulation scripts and notebooks referenced in the supplementary
document are shipped alongside the software release described in
Appendix~\ref{sec:software}.

\section{Software availability}\label{sec:software}

A reference Python~3 implementation of CRTB, together with the
sparse variant, the cellwise pre-filter and the SPADIMO-based
diagnostics, will be made available as part of release
\texttt{0.5} of the open-source \texttt{twoblock} package
(\url{https://pypi.org/project/twoblock/}), alongside the
scripts that reproduce the simulation study and the two real-data
examples reported in this paper.

\bibliographystyle{apalike}
\bibliography{crtb_paper}

\end{document}